\documentclass[letterpaper,english,reprint, aps]{revtex4-1}
\usepackage[T1]{fontenc}
\usepackage[latin9]{inputenc}
\usepackage{babel}
\usepackage{amsmath}
\usepackage{amssymb}
\usepackage{graphicx}
\usepackage[unicode=true]
 {hyperref}

\makeatletter

\newcommand*\LyXbar{\rule[0.585ex]{1.2em}{0.25pt}}

\providecommand{\tabularnewline}{\\}

\makeatother

\begin{document}
\title{Percolation properties of the classic Sierpinski carpet and sponge}
\author{Clinton DeW.\,Van Siclen}
\email{cvansiclen@gmail.com}

\address{1435\,W 8750\,N, Tetonia, Idaho 83452, USA}
\date{19 February 2023}
\begin{abstract}
Iterative construction of a Sierpinski carpet or sponge is shown to
be a critical phenomenon analogous to uncorrelated percolation. Critical
exponents are derived or calculated (by random walks over the carpet
or sponge at infinite iteration) that are related by equations identical
to those obtained from percolation theory. Finite-size scaling then
gives accurate values for the scalar transport properties (e.g., effective
conductivity) of the carpet or sponge at any stage of iteration.
\end{abstract}
\maketitle

\section{Introduction}

The classic Sierpinski carpet \citep{Mandelbrot} is a recursive,
self-similar fractal embedded in two-dimensional (2D) Euclidean space.
The generator is shown in Fig.\,1 (the center square is removed,
leaving the surrounding eight squares); the carpet after the third
iteration is shown in Fig.\,2. This center-hole $(3,1)$ Sierpinski
carpet has Hausdorff (fractal) dimension
\begin{equation}
\mathcal{H}=\frac{\ln\left(b^{2}-m\right)}{\ln b}=\frac{\ln8}{\ln3}\approx1.89279\label{eq:1}
\end{equation}
given the scaling factor $b=3$ and number $m=1$ of eliminated squares
in the generator.

\begin{figure}[b]
\includegraphics[scale=0.5]{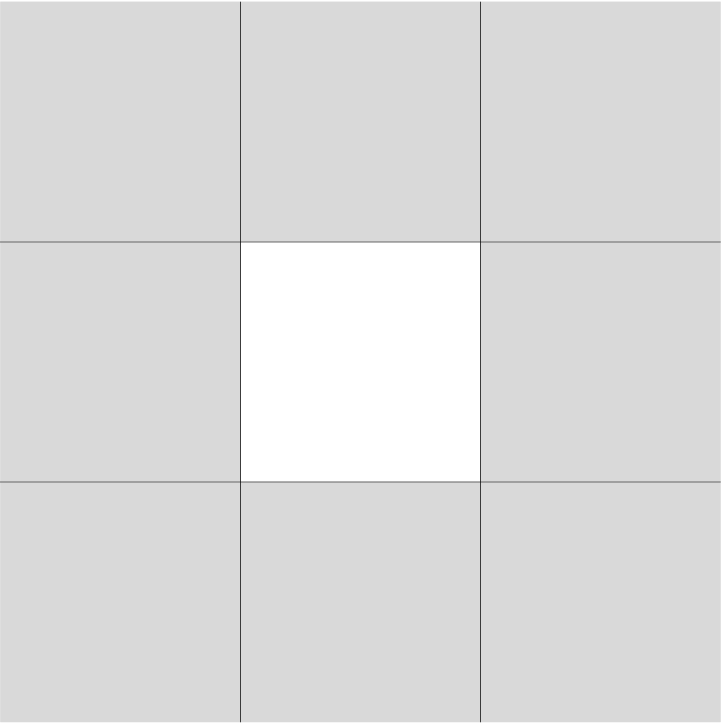}

FIG. 1. Generator for the center-hole $\left(3,1\right)$ Sierpinski
carpet.
\end{figure}

\begin{figure}[b]
\includegraphics[scale=0.5]{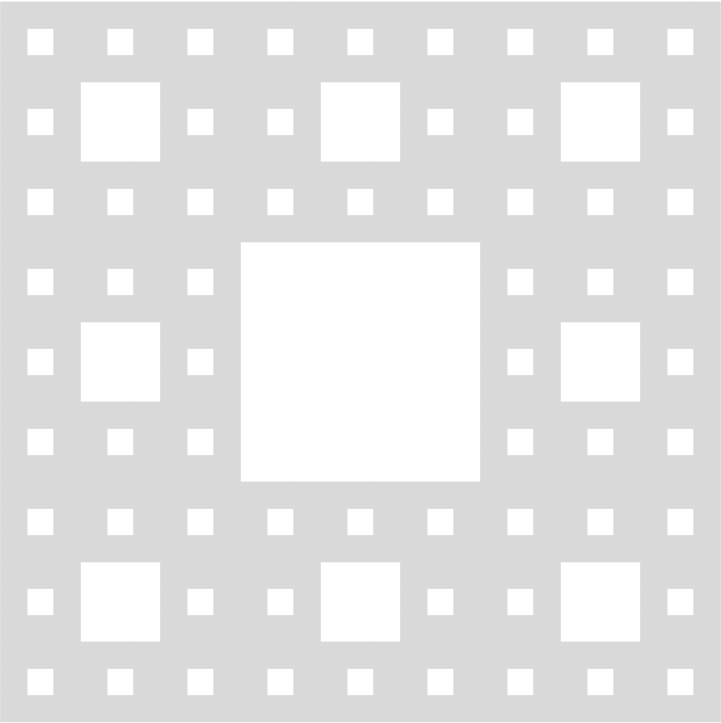}

FIG. 2. Sierpinski carpet at iteration $i=3$.
\end{figure}

With each iteration one-ninth of the mass is removed while the carpet
remains self-similar. The correlation length $\xi$ for this system
is the length of the carpet, and so increases to infinity\,\LyXbar \,when
measured by comparison to the size\textit{ $\mathrm{h}_{i}$} of the
smallest hole\,\LyXbar \,as the number \textit{i} of iterations
goes to infinity. Thus the iterative construction of this recursive,
self-similar fractal resembles the approach to a percolation threshold.

In this paper, relations between critical exponents are derived that
are identical to those characterizing uncorrelated percolation. This
suggests the term ``recursion percolation'' for the critical phenomena
exhibited by recursive fractals.

The following section briefly describes the Walker Diffusion Method
(WDM) by which the analytical and numerical results for the 2D Sierpinski
carpet and 3D Sierpinski sponge are obtained. The percolation analogy
is then developed in Sec.\,III. Values for parameters specific to
the transport properties of the carpet and sponge cannot be derived:
those are calculated in Sec.\,IV by random walks over a carpet or
sponge of infinite size (infinite iteration). Relevant papers in the
literature are discussed in Sec.\,V. Concluding remarks are made
in Sec.\,VI.

\section{Walker Diffusion Method}

This application of the WDM \citep{CVS99} utilizes the relation
\begin{equation}
\sigma=\left\langle \sigma(\mathbf{r})\right\rangle D_{w}\label{eq:2}
\end{equation}
between the effective conductivity $\sigma$ of a composite material
and the (dimensionless) diffusion coefficient $D_{w}$ obtained from
walkers diffusing through a digital representation of the composite.
The factor $\left\langle \sigma(\mathbf{r})\right\rangle $ is the
volume average of the constituent conductivities.

In effect, the phase domains that make up the composite are host to
walker populations, where the walker density of a population is proportional
to the conductivity value of its host domain. The principle of detailed
balance ensures that the population densities are maintained, by providing
the following rule for walker diffusion over the digitized composite:
a walker at site (or pixel/voxel) \textit{i} attempts a move to a
randomly chosen adjacent site \textit{j} during the time interval
$\tau=\left(4d\right)^{-1}$, where \textit{d} is the Euclidean dimension
of the system; this move is successful with probability $p_{ij}=\sigma_{j}/\left(\sigma_{i}+\sigma_{j}\right)$,
where $\sigma_{i}$ and $\sigma_{j}$ are the conductivities of sites
\textit{i} and \textit{j}, respectively. The path of the walker thus
reflects the composition and morphology of the domains that are encountered.

The diffusion coefficient $D_{w}$ is calculated using the equation
\begin{equation}
D_{w}=\frac{\left\langle R(t)^{2}\right\rangle }{2\,d\,t}\label{eq:3}
\end{equation}
where the set $\left\{ R\right\} $ of walker displacements, each
occurring over the time interval $t$, must have a Gaussian probability
distribution that is necessarily centered well beyond $\xi$. The
correlation length $\xi$ is identified as the length scale above
which a composite material attains the ``effective'', or macroscopic,
value of a scalar transport property (electrical conductivity, for
example).

For displacements $R<\xi$, the walker diffusion is \textit{anomalous}
rather than Gaussian due to the heterogeneity of the composite at
length scales less than $\xi$. However there is an additional characteristic
length $\xi_{0}<\xi$, below which the composite is \textit{effectively}
homogeneous. Then a walker displacement of $\xi$ requiring a travel
time $t_{\xi}=\xi^{2}/\left(2\,d\,D_{w}\right)$ is produced by a
walk comprising $\left(\xi/\xi_{0}\right)^{d_{w}}$ segments of length
$\xi_{0}$, each requiring a travel time of $t_{0}=\xi_{0}^{2}/\left(2\,d\,D_{0}\right)$,
where $D_{0}$ is the walker diffusion coefficient calculated from
displacements $R\leq\xi_{0}$. Thus $t_{\xi}=\left(\xi/\xi_{0}\right)^{d_{w}}t_{0}$,
which gives the relation
\begin{equation}
D_{w}=D_{0}\left(\frac{\xi}{\xi_{0}}\right)^{2-d_{w}}\label{eq:4}
\end{equation}
between the walker diffusion coefficient $D_{w},$ the fractal dimension
$d_{w}$ of the walker path, and the correlation length $\xi$.

\section{Recursion percolation}

For convenience the Sierpinski carpet or sponge at iteration \textit{i}
is denoted by $\mathrm{S}_{i}$. Further, all lengths, areas, and
volumes are in units of the smallest feature, which is the smallest
hole $\mathrm{h}_{i}$. That is necessarily the size of a single pixel
or voxel, so $\mathrm{h}_{i}$ has length/area/volume equal to 1.

As these fractals are self-similar at \textit{all} length scales,
the characteristic length $\xi_{0}=h_{i}=1$.

With each iteration \textit{i}, the areal fraction $a_{i}$ of the
$\mathrm{S}_{i}$ carpet that is conducting (i.e., not lost to cut-outs)
decreases according to $a_{i}=\left(8/9\right)^{i}$, while the correlation
length $\xi^{(i)}$ increases according to $\xi^{(i)}=3^{i}\,\xi_{0}=3^{i}$.
These two equalities can be written $\ln a_{i}=i\ln\left(8/9\right)$
and $\ln\xi^{(i)}=i\ln3$, respectively, so producing the power-law
relation
\begin{equation}
\xi^{(i)}=a_{i}^{-\nu}\label{eq:5}
\end{equation}
with the exponent
\begin{equation}
\nu=\frac{\ln3}{\ln\left(9/8\right)}=\left(2-\mathcal{H}\right)^{-1}.\label{eq:6}
\end{equation}
This result makes the case that recursive construction of the Sierpinski
carpet is a percolation-like phenomenon \citep{S=000026A}. Of course,
the ``percolation threshold'' is approached (from above) as iteration
$i\rightarrow\infty$, causing $a_{i}\rightarrow0$.

According to Eqs.\,(\ref{eq:2}) and (\ref{eq:4}), the effective
conductivity of $\mathrm{S}_{i}$ is
\begin{equation}
\sigma^{(i)}=\sigma_{1}\,a_{i}\,D_{w}^{(i)}=\sigma_{1}\,a_{i}\,D_{0}\left(\frac{\xi^{(i)}}{\xi_{0}}\right)^{2-d_{w}}\label{eq:7}
\end{equation}
where $a_{i}$ is the areal fraction of the carpet that is conducting,
and $\sigma_{1}$ is the conductivity of that material. The walker
diffusion coefficient $D_{0}<1$ because walkers near the non-conducting
cut-outs tend to linger there (an effect of the walker diffusion rule
stated in Sec.\,II). Both $D_{0}$ and the walker path dimension
$d_{w}$, which pertain to the transport properties of the carpet,
must be calculated, not derived.

Use of Eq.\,(\ref{eq:5}) in Eq.\,(\ref{eq:7}) gives

\begin{equation}
\sigma^{(i)}=\sigma_{1}\,a_{i}\,D_{0}\left(\xi^{(i)}\right)^{2-d_{w}}=\sigma_{1}\,D_{0}\,a_{i}^{t}\label{eq:8}
\end{equation}
with the exponent
\begin{equation}
t=1+\nu\left(d_{w}-2\right).\label{eq:9}
\end{equation}
Then use of Eq.\,(\ref{eq:5}) in Eq.\,(\ref{eq:8}) produces the
\textit{asymptotic} relation
\begin{equation}
\sigma(\xi)\sim\sigma_{1}\,D_{0}\,\xi^{-t/\nu}\label{eq:10}
\end{equation}
giving the finite-size scaling relation
\begin{equation}
\sigma(L)=\sigma_{1}\,D_{0}\,L^{-t/\nu}\label{eq:11}
\end{equation}
for all $L=3^{i}$. {[}Note that $\sigma(L)$ is the effective conductivity
of an infinite 2D array of carpets of size $L$. The length $L$ plays
the role of the correlation length $\xi$.{]}

The formalism above is straightforwardly applied to the Sierpinski
sponge as well. All relations for the sponge are obtained by replacing
$a_{i}$ with $v_{i}$ in the equations above. Here $v_{i}=\left(26/27\right)^{i}$
is the volume fraction of the $\mathrm{S}_{i}$ sponge that is conducting.
The Hausdorff dimension of the sponge is
\begin{equation}
\mathcal{H}=\frac{\ln\left(b^{3}-m\right)}{\ln b}=\frac{\ln26}{\ln3}\approx2.96565\label{eq:12}
\end{equation}
giving the exponent
\begin{equation}
\nu=\frac{\ln3}{\ln\left(27/26\right)}=\left(3-\mathcal{H}\right)^{-1}.\label{eq:13}
\end{equation}

As the exponent $d_{w}$ is central to the development above, it is
interesting to consider whether that development imposes any limits
on its value. In fact an analytic bound is obtained by comparing the
\textit{asymptotic} behavior (meaning: as iteration $i\rightarrow\infty$)
of $D_{w}^{(i)}$ with that of the conducting areal fraction $a_{i}$
of the carpet or conducting volume fraction $v_{i}$ of the sponge.
From Eq.\,(\ref{eq:4}),

\begin{equation}
D_{w}(\xi)\sim\xi^{2-d_{w}}\label{eq:13a}
\end{equation}
and from Eq.\,(\ref{eq:5}),
\begin{equation}
a(\xi)\sim\xi^{-1/\nu}.\label{eq:13b}
\end{equation}
The value $D_{w}^{(i)}$, reflecting the walker behavior, is responsive
to the value $a_{i}$ or $v_{i}$ (rather than vice versa), suggesting
that $1/\nu>d_{w}-2$. Thus

\begin{equation}
d_{w}<2+\frac{1}{\nu}=2+(d-\mathcal{H}).\label{eq:13c}
\end{equation}
Then $d_{w}<2.10721$ in the case of the 2D carpet, and $d_{w}<2.03435$
in the case of the 3D sponge.

Note that Eq.\,(\ref{eq:13c}), giving an analytic upper bound for
$d_{w}$, should apply to similar recursive fractals as well.

The relations in this section are recognizable from standard percolation
theory \citep{CVSperc}. In particular, the latter gives $\nu/\beta=\left(d-D\right)^{-1}$
and $t=\beta+\nu\left(d_{w}^{*}-2\right)$, where $D$ is the fractal
dimension of the incipient infinite cluster of conducting sites, and
the exponent $\beta$ is less than 1, reflecting the fact that conductor
sites not belonging to the percolating cluster cannot contribute to
the conductivity of the system. In contrast, the Sierpinski carpet
and sponge do not have such ``stranded'' conductor sites (thus the
``percolation thresholds'' $a_{\infty}=0$ and $v_{\infty}=0$).

A more important distinction is that in standard percolation the exponent
relations and values are obtained in the limit $\xi\rightarrow\infty$;
that is, \textit{at} the percolation threshold. Because $d_{w}$ increases
from $2$ to $d_{w}^{*}$ as the correlation length $\xi\rightarrow\infty$,
the derivation that leads to Eq.\,(\ref{eq:13c}) is not applicable:
that derivation relies on the constancy of the $d_{w}$ value over
all iterations of the carpet and sponge.

\section{Numerical methods and results}

The value of the walker path dimension $d_{w}$, and the value of
the diffusion coefficient $D_{0}$ associated with the length scale
$\xi_{0}$, must be calculated. For a fractal system of finite size
$L$, Eq.\,(\ref{eq:4}) may be written
\begin{equation}
D_{w}(L)=D_{0}L^{2-d_{w}}.\label{eq:14}
\end{equation}
This relation can be expressed in terms of the computable variable
$\left\langle R(t)^{2}\right\rangle $:
\begin{equation}
\frac{\left\langle R(t)^{2}\right\rangle }{2\,d\,t}=D_{0}\left\langle R(t)^{2}\right\rangle ^{1-d_{w}/2}\label{eq:15}
\end{equation}
which simplifies to
\begin{equation}
\left\langle R(t)^{2}\right\rangle =\left(2\,d\,tD_{0}\right)^{2/d_{w}}.\label{eq:16}
\end{equation}
Thus walks over the fractal system will produce points $(\ln t,\ln\left\langle R(t)^{2}\right\rangle )$
that satisfy the equation
\begin{equation}
\ln\left\langle R(t)^{2}\right\rangle =\frac{2}{d_{w}}\ln t+\frac{2}{d_{w}}\ln\left(2\,dD_{0}\right).\label{eq:17}
\end{equation}
A linear fit to the points produces a plot from which the values $d_{w}$
and $D_{0}$ can be ascertained.

This graphical approach is taken for the Sierpinski carpet and sponge.
These may be created using the subroutine given in Appendix A. Note
that in both cases the smallest hole, being indivisible, is the size
of a single site (pixel or voxel).

Walks over the carpet or sponge are accomplished by use of the variable
residence time algorithm \citep{CVS99}, described in Appendix B.
The algorithm takes advantage of the statistical nature of the diffusion
process to eliminate (while accounting for) unsuccessful attempts
by the walker to move to a neighboring site.

To allow very long walks, all walks are actually taken over a carpet
or sponge at \textit{infinite} iteration. Note that the subroutine
locates conducting and non-conducting sites with respect to an origin,
which in the case of the sponge is the $\left(i,j,k\right)$ site
with coordinates $\left(0,0,0\right)$. Thus the sponge occupies all
space with site index values $i,j,k$ greater than or equal to $0$.
A move by a walker at $\left(i,j,k\right)$ is of course determined
by the conductivities of the adjacent sites: those values (1 or 0)
are obtained by calls to the subroutine.

Figure 3 is the plot of points obtained from walks over the infinite
Sierpinski carpet. The slope $2/d_{w}$ of the fitted line gives the
value $d_{w}=2.09675$. The y-intercept $(2/d_{w})\ln(2\,d\,D_{0})$
of the line gives the value $D_{0}=0.77376$.

\begin{figure}
\includegraphics[scale=0.8]{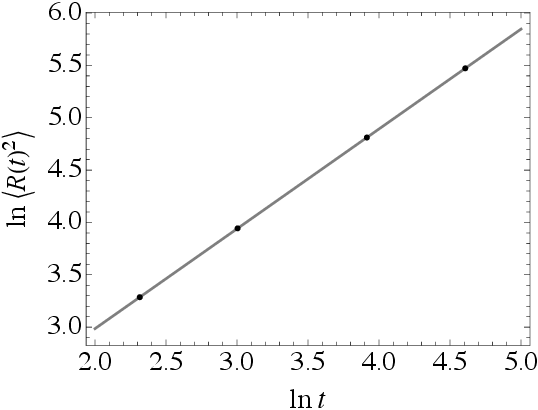}

FIG. 3. Linear fit to data points obtained from walks over the Sierpinski
carpet.
\end{figure}

Each point in Fig.\,3 is obtained from $40$ sequences of $10^{6}$
walks of time $t$. {[}A \textit{sequence} of $10^{6}$ walks is actually
a single walk of time $10^{6}\times t$. During that long walk every
displacement $R(t)$ is recorded, for a total of $10^{6}$ displacements.{]}
The plotted value $\left\langle R(t)^{2}\right\rangle $ is the average
of \textit{all} walks of time $t$ (that is, the average of all sequences).
In every case the number of sequences is sufficient that additional
sequences would change the value $\left\langle R(t)^{2}\right\rangle $
by only an insignificant amount (far less than the point size in the
figure).

A sequence of walks is initiated by placing a walker at a randomly
chosen conducting site $\left(i\gg0,j\gg0\right)$ of the infinite
carpet.

Similar calculations are made for the 3D Sierpinski sponge.

\begin{figure}
\includegraphics[scale=0.8]{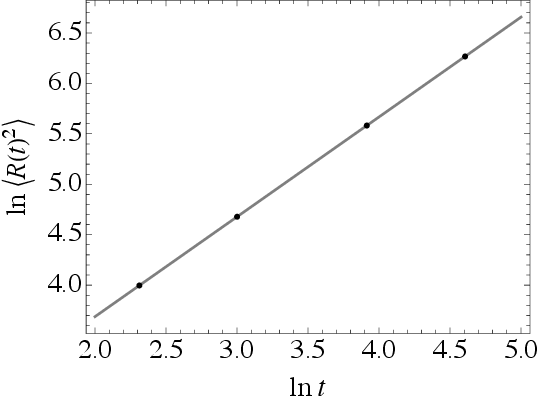}

FIG. 4. Linear fit to data points obtained from walks over the Sierpinski
sponge.
\end{figure}

Figure 4 is the plot of points obtained from walks over the infinite
Sierpinski sponge. The slope $2/d_{w}$ of the fitted line gives the
value $d_{w}=2.02026$. The y-intercept $(2/d_{w})\ln(2\,d\,D_{0})$
of the line gives the value $D_{0}=0.935312$.

Each point in Fig.\,4 is obtained from $40$ sequences of $10^{6}$
walks of time $t$.

For convenience the numerical results obtained above are presented
together in Table I.

\begin{table}[h]
TABLE I. Calculated values.

\medskip{}

\begin{tabular}{|c|c|c|c|}
\hline 
$d$ & $d_{w}$ & $D_{0}$ & $t/\nu$\tabularnewline
\hline 
\hline 
$2$ & $2.09675$ & $0.77376$ & $0.203957$\tabularnewline
\hline 
~$3$~ & ~$2.02026$~ & ~$0.935312$~ & ~$0.0546151$~\tabularnewline
\hline 
\end{tabular}
\end{table}

Interestingly, an upper bound on the value $d_{w}$ can be obtained
by considering the number of steps in a walk. Recall from Sec.\,II
that a walker displacement $\xi$ is produced by $(\xi/\xi_{0})^{d_{w}}$
steps, each of length $\xi_{0}$. Thus the relation between the number
of steps $n=(\xi/\xi_{0})^{d_{w}}$ and displacement $\xi$ is
\begin{equation}
\frac{\xi}{\xi_{0}}=n^{1/d_{w}}.\label{eq:17a}
\end{equation}

This relation applies to displacements $R<\xi$ as well. However,
calculations for short walks may be affected by the finite size of
the sites that compose the carpet and sponge. Therefore consider the
relation
\begin{equation}
\frac{R}{\xi_{0}}=n^{1/\delta}\label{eq:17b}
\end{equation}
where both the displacement $R$ and the exponent $\delta$ are determined
by the number of steps $n$. In any case, the calculated value $\delta\rightarrow d_{w}$
as $R\rightarrow\xi$.

These Sierpinski fractals have the characteristic length $\xi_{0}=1$.
Then it is computationally convenient to square both sides of Eq.\,(\ref{eq:17b}),
since $R^{2}$ has an integer value. With these changes, Eq.\,(\ref{eq:17b})
produces the relation
\begin{equation}
\delta(n)=\frac{2\,\ln n}{\ln\left\langle R(n)^{2}\right\rangle }\label{eq:17c}
\end{equation}
where $\left\langle R(n)^{2}\right\rangle $ is the average of all
the $R^{2}$ values obtained from a very large number of walks of
$n$ steps.

Thus the exponent $\delta(n)\rightarrow d_{w}$ as the number $n$
of steps in a walk increases. This is apparent in Table II, where
values $\delta(n)>d_{w}$ are recorded. Each value is obtained from
$40$ sequences of walks, each sequence comprising $10^{5}$ walks,
over the \textit{infinite} carpet or sponge. The average value $\left\langle R(n)^{2}\right\rangle $
used in Eq.\,(\ref{eq:17c}) is taken from \textit{all} walks of
$n$ steps.

\begin{table}[h]
TABLE II. Exponent $\delta(n)$ values for $R<\xi$.

\medskip{}

\begin{tabular}{|c|c|c|}
\hline 
~steps/walk~ & $\delta_{\textrm{2D}}$ & $\delta_{\textrm{3D}}$\tabularnewline
\hline 
\hline 
$10^{2}$ & ~$2.11382(175)$~ & ~$2.02364(107)$~~\tabularnewline
\hline 
$10^{3}$ & $2.10737(103)$ & $2.02219(89)$\tabularnewline
\hline 
$10^{4}$ & $2.10526(101)$ & $2.02150(60)$\tabularnewline
\hline 
$10^{5}$ & $2.10357(82)$ & $2.02127(41)$\tabularnewline
\hline 
\end{tabular}
\end{table}

Note that a value given as $1.234(5)$ means $1.234$ with standard
deviation $0.005$, and so indicates the range $1.229$ to $1.239$,
centered on $1.234$. The standard deviation for $\delta(n)$ is calculated
from the $40$ values obtained by the $40$ sequences of walks.

\section{Prior research}

A number of papers report calculations (and in one case an experiment)
that directly or indirectly produce a value for $d_{w}$. Note that
a value given as $1.234(5)$ indicates the range $1.229$ to $1.239$,
centered on $1.234$.

Gefen \textit{et al}.\,\citep{Gefen} overlay a resistor network
on the classic Sierpinski carpet, and obtain a finite-size scaling
relation for the resistance by a renormalization method. That produces
the resistance exponent $\widetilde{\zeta}_{R}=0.194$, giving $d_{w}=\mathcal{H}+\widetilde{\zeta}_{R}=2.087$.
(Note, by the way, that $t/\nu=d-2+\widetilde{\zeta}_{R}$ \citep{S=000026A}.)

Barlow \textit{et al}.\,\citep{Barlow} consider resistor network
approximations of the classic Sierpinski carpet, and apply electrical
circuit theory to calculate the spectral dimension $d_{s}=1.80525$,
giving $d_{w}=2\,\mathcal{H}/d_{s}=2.09698$.

Kim \textit{et al}.\,\citep{Kim} perform random walks on the $\mathrm{S}_{7}$
carpet, and obtain $d_{w}=2.106(16)$ from their relation $\left\langle R(N)^{2}\right\rangle \propto N^{2/d_{w}}$,
where $N$ (with values up to $10^{4}$) is the number of steps in
a walk.

Zhuang \textit{et al}.\,\citep{Zhuang} measure the resistivity for
carpets $\mathrm{S}_{1}$, $\mathrm{S}_{2}$, and $\mathrm{S}_{3}$
cut out of copper and aluminum sheets. Those data points lie on a
line $\ln\left(1/R\right)\propto\ln L$, where $R(L)$ is the resistance
and $L$ is the size scaling of the samples, indicating finite-size
scaling with conductivity exponent $t/\nu=0.22(1)$.

Aar$\tilde{\mathrm{a}}$o Reis \citep{AaraoReis} performs random
walks on carpets $\mathrm{S}_{2}$, $\mathrm{S}_{3}$, $\mathrm{S}_{4}$,
and $\mathrm{S}_{5}$, and overlays the four plots of $\left\langle R(N)^{2}\right\rangle ^{1/2}/L$
versus $L\,N^{-\nu_{w}}$ ($L=3^{i}$, and $N$ up to $8\times10^{4}$
steps). The best data collapse (coincidence of the four curves) occurs
for $\nu_{w}=0.476(5)$, giving $d_{w}=1/\nu_{w}=2.101(22)$. This
approach is taken as well for sponges $\mathrm{S}_{2}$, $\mathrm{S}_{3}$,
and $\mathrm{S}_{4}$, producing $\nu_{w}=0.492(6)$, giving $d_{w}=1/\nu_{w}=2.033(25)$.

Suwannasen \textit{et al}.\,\citep{Suwannasen} perform random walks
with $2^{15}$ walkers initially distributed at random on the $\mathrm{S}_{15}$
carpet. Their log-log plot of $\left\langle R(t)^{2}\right\rangle $
versus $C\,t^{2/d_{w}}$ gives $d_{w}=2.10(1)$.

\section{Concluding remarks}

A notable aspect of this paper is the appearance of the factor $D_{0}$.
Most importantly, it appears in the finite-size scaling relation Eq.\,(\ref{eq:11})
giving the effective conductivity of a finite carpet or sponge.

Random walk calculations that relate the number of steps in a walk
to the consequent walker displacement $R$ are shown to give an upper
bound for the value of the exponent $d_{w}$. The longer the walk
(that is, as $R\rightarrow\xi$), the tighter the upper bound.

In addition, an analytic expression giving an upper bound for $d_{w}$
is derived by comparing the asymptotic ($i\rightarrow\infty$) behavior
of $D_{w}^{(i)}$, and $a_{i}$ or $v_{i}$, governed by the exponents
$d_{w}$ and $\nu$, respectively. Thus Eq.\,(\ref{eq:13c}) should
apply to similar recursive fractals as well.

The Sierpinski sponge has been used as a heuristic model for porous
rock, which is typically found to have a fractal character over some
range of length scales. In fact the relations derived here carry over
to isotropic porous rock, with the exception that the conducting volume
fraction $v_{i}$ of the $\mathrm{S}_{i}$ sponge is replaced by the
connected porosity $\phi$ of the rock, given by the relation
\begin{equation}
\phi=\frac{\left(\frac{\xi}{\xi_{0}}\right)^{D}}{\left(\frac{\xi}{\xi_{0}}\right)^{3}}=\left(\frac{\xi}{\xi_{0}}\right)^{D-3}\label{eq:18}
\end{equation}
where the right-hand side of the equality is the volume fraction of
the rock that is occupied by the connected pore space of (fractal)
mass dimension $D<3$. The corresponding exponent is $\nu=(3-D)^{-1}$.

A porous rock sample of size $L$ saturated with an electrolyte solution
having conductivity $\sigma_{e}$ would exhibit an effective conductivity
given by the finite-size scaling relation Eq.\,(\ref{eq:11}):
\begin{equation}
\sigma(L)=\sigma_{e}\,D_{0}\left(\frac{L}{\xi_{0}}\right)^{-t/\nu}=\sigma_{e}\,\phi\,D_{0}\left(\frac{L}{\xi_{0}}\right)^{2-d_{w}}.\label{eq:19}
\end{equation}
\medskip{}

\begin{acknowledgments}
I thank Professor Indrajit Charit (Department of Nuclear Engineering
\& Industrial Management) for arranging my access to the resources
of the University of Idaho Library (Moscow, Idaho).
\end{acknowledgments}

\appendix

\subsection*{Appendix A: Sierpinski fractal construction subroutine}

This subroutine determines whether element $i,j$ (carpet) or $i,j,k$
(sponge) of the array representing the Sierpinski fractal is conducting
or insulating, and returns the value $1$ or $0$, respectively. Note
that a corner of the array is an element with $i=j=k=0$. This implementation
is written in C.

\medskip{}

\noindent int SierpinskiFractal(int i, int j, int k)

\noindent \{

\noindent \enskip{}while (i>0 || j>0 || k>0) \{

\enskip{}if (i\%3 == 1 \&\& j\%3 == 1 \&\& k\%3 == 1) return 0;

\enskip{}i /= 3;

\enskip{}j /= 3;

\enskip{}k /= 3;

\noindent \enskip{}\}

\noindent \enskip{}return 1;

\noindent \}

\subsection*{Appendix B: Variable residence time algorithm}

According to this algorithm \citep{CVS99}, the actual behavior of
the walker is well approximated by a sequence of moves in which the
direction of the move from a site $i$ is determined randomly by the
set of probabilities $\left\{ P_{ij}\right\} $, where $P_{ij}$ is
the probability that the move is to adjacent site $j$ (which has
conductivity $\sigma_{j}$) and is given by the equation
\[
P_{ij}=\frac{\sigma_{j}}{\sigma_{i}+\sigma_{j}}\left[\sum_{k=1}^{2d}\left(\frac{\sigma_{k}}{\sigma_{i}+\sigma_{k}}\right)\right]^{-1}.\tag{{B1}}
\]
The sum is over all sites adjacent to site $i$. The time interval
over which the move occurs is
\[
T_{i}=\left[2\sum_{k=1}^{2d}\left(\frac{\sigma_{k}}{\sigma_{i}+\sigma_{k}}\right)\right]^{-1}.\tag{{B2}}
\]
Note that this version of the variable residence time algorithm is
intended for orthogonal systems (meaning a site in a 3D system has
six neighbors, for example).

\end{document}